\documentstyle[aps,twocolumn,epsf]{revtex} %%\tightenlines

\begin{document} 
\draft

\twocolumn[\hsize\textwidth\columnwidth\hsize\csname
@twocolumnfalse\endcsname

\tightenlines

\title{Effective interaction between helical bio-molecules}

\author{E.Allahyarov
 $^{1,2}$,\,\, H.L\"owen $^{1}$\,\, }

\address{{1} Institut\, f\"{u}r\, Theoretische \,Physik
  II,\,Heinrich-Heine-Universit\"{a}t\,
 D\"{u}sseldorf,\,\mbox{D-40225}\,D\"{u}sseldorf, \,Germany}
\address{{2} Institute\, for\, High\, Temperatures,\, Russian \,Academy \,of \,Sciences,\, 127412
\,Moscow,\, Russia }

\date{\today}

\maketitle
\begin{abstract}
The effective interaction between two
parallel strands of helical bio-molecules, such as deoxyribose nucleic
acids (DNA), is calculated using computer simulations of the
``primitive" model of electrolytes. 
In particular we study a simple model for B-DNA incorporating
explicitly its charge pattern as a double-helix structure. The
effective force and the effective torque exerted onto the molecules
depend on the central distance and on the relative  orientation.
The contributions of nonlinear screening by monovalent counterions
to these forces and torques are analyzed and calculated 
for different salt concentrations.
As a result, we find that the sign of the force depends 
sensitively on the relative orientation. For intermolecular 
distances smaller than $6\AA$
 it can be both attractive and repulsive.
Furthermore we report a nonmonotonic behaviour of the 
effective force for increasing salt
concentration. Both features cannot be described within 
linear screening theories.
For  large  distances, on the other hand, the
results agree with linear screening theories provided 
the  charge of the bio-molecules is suitably renormalized.

\end{abstract}

\vskip2pc]

\renewcommand{\thepage}{\hskip 8.9cm \arabic{page} \hfill Typeset
using REV\TeX }
\narrowtext

\section{Introduction}

Aqueous solutions of helical bio-molecules 
like deoxyribose nucleic acids (DNA) %\cite{Saengerbook}
 are typically highly charged
such that electrostatic
interactions  play an important role in many aspects of their
structure and function \cite{jaya,zimm,lin,lebret2,klein,bloomfield}. 
Understanding  the total effective
interaction between two helical molecules 
is  important since  this  governs the self-assembly of bio-molecules,
like bundle formation and DNA condensation or compaction which in turn
is fundamental for gene delivery and gene therapy.
In aqueous solution,
such rod-like  polyelectrolytes release counterions in the solution 
which ensure  global charge neutrality of the system.
Together with these counterions, there are, in general, added salt ions 
dissolved in the solution. The  thermal
ions screen the bare electrostatic
interactions between the bio-molecules, such that the effective interaction 
between them is expected to become weaker than the direct Coulomb repulsion. 
For very high concentrations of bio-molecules or short distances even
a mutual attraction due to counterion "overscreening"  is conceivable
  \cite{gronbech,nilsson,guldbrand,bolhius,allah,pincus,tang,mashl,rouzina,wilson1,widom,kjellander,kekicheff,kepler,khan,namlee,ueda,lyubartsev}.

In this paper, we study the effective interaction between two parallel helical
bio-molecules. In particular, we investigate how the electrostatic
interactions are influenced by details of the charge pattern on the biological
macromolecules. In fact, in many cases, as e.g.\ for DNA molecules,
the charge pattern on the molecules is not uniform but exhibits
an intrinsic helix structure. If two parallel helical
molecules are nearby, this helix structure
will induce an interaction that depends on the relative orientation
of the two helices. Our studies are based on  computer simulation of the ``primitive"
 model of
electrolytes \cite{Hansen}. In particular we study a simple model for B-DNA\@. 
This model explicitly takes into account  the double-helical charge
pattern along the DNA-strand, it also accounts for the molecular shape
by modeling the major and minor grooves along the strand.
The charged counter- and salt ions in the solutions are explicitly 
incorporated into our model. On the other hand,  the water molecules 
only constitute a continuous background with a dielectric constant $\epsilon$
screening the Coulomb interactions. Hence the discrete nature of the solvent is neglected
as well as more subtle effects as image charges induced by dielectric discontinuities
at the DNA-water boundary \cite{jaya2,stigter,troll,kar}, 
hydration effects due to the affection of the hydrophilic surface to
the interfacial layers of water 
\cite{parsegian,leikinpar1,rand,rau,gruen,mariani}, and
spatial dependent dielectric constants resulting from the
decreasing water  mobility in confining geometries and from saturation effects
induced by water polarization  near the highly charged molecular
surfaces \cite{mazur,hingerty,jaya1,lamm1,luka1,maarel}.

Our motivation to consider such a simple "primitive"  model is threefold:
First, though solvent effects seem to be relevant they should average out on a
length scale which is larger than the range of the microscopic sizes.
Hence the electrostatic effects are expected to
 dominate the total effective interactions.
Second, it is justified to  study a simple model completely and then adapt
it by introducing more degrees of freedom in order to better match the 
experimental situation. Our philosophy is indeed to  understand the
principles of a simple model first and then turn step by step to more complicated
models. Third, even within the ``primitive" approach, there are many unsolved problems
and unexpected effects such as mutual attraction of equally charged particles.
Our computer simulation method has the advantage that ``exact" results  are obtained
that reflect directly the nature of the model. Hence we get rid of any approximation
inherent in a theoretical description. Consequently, the dependence
of the effective interactions on a model parameter can systematically be 
studied and the trends can be compared to experiments. In this respect our model
is superior to previous studies that describe the counterion screening 
by linear Debye-H{\"u}ckel  \cite{lamm1,fogo,lebret2,wagner}
or nonlinear Poisson-Boltzmann theory \cite{jaya2,lebret2,pack,mills,gavryushov,murthy,vlachy,paulsen,pack2,granot}
and even to recent approaches that include approximatively counterion correlations
\cite{shklovskii,levin}.
We also emphasize that one main goal of the paper is to incorporate the
molecular shape and charge pattern explicitly which is modelled in many studies
simply as a homogeneously charged cylinder \cite{lamm1,lebret2,luka,conrad}.
In fact we find that the double-helix structure has an important
influence on the effective interaction for surface-to-surface separations smaller
than  $6\AA$. 
In detail, the interaction  can be both repulsive and attractive depending
on the relative orientation and the mutual distance between two parallel
DNA strands. This effect which is typically ignored in the charged-cylinder model for DNA
 will significantly affect the self-assembly of parallel smectic layers of
DNA fragments and may result in unusual crystalline structures at high
concentrations.

Let us also mention that many theoretical studies involve only a single DNA molecule 
\cite{jaya3,hochberg,edwards,hoch2,lin}. To extract the effective interaction, 
however, one has at least to include two molecules in the model which is the
purpose of the present paper.
In this study we only consider monovalent counterions.
Multivalent counterions and a more detailed survey on the
influence  of model parameters on the effective interactions
will be considered in a subsequent publication.

The remainder of paper is organized as follows. In chapter II,
 we present the details of the model used in this paper. Chapter
III describes the target quantities of the applied model. Simulation
 details are presented in chapter IV\@.
Theories based on linear screening approaches such as the
homogeneously charged cylinder model, the Yukawa segment model and the
Kornyshev-Leikin theory \cite{kor1}   are shortly discussed 
in chapter V.  Results of the simulation and their
comparison to  linear screening theories  are contained
in sections VI-VIII for the point-charge model, the grooved model and added
salt respectively.  We conclude
in section IX.

\section{The model}
\label{model}

\begin{figure}
   \epsfxsize=7cm %9cm
   \epsfysize=9cm %12cm
~\hfill\epsfbox{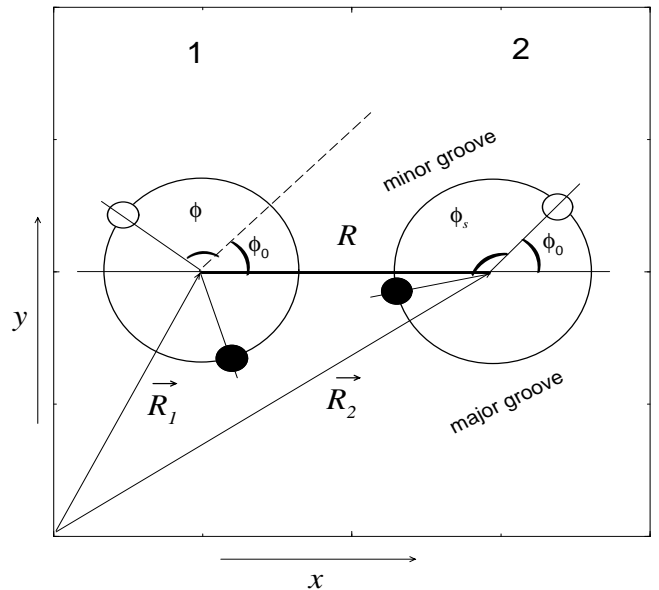}\hfill~
   \caption{A schematic picture explaining the positions of
     DNA molecules and the definition of the different azimuthal
     angles $\phi_0, \phi, \phi_s$. For further information see text.}
   \label{xyplane}
\end{figure}

The charge pattern and the shape of
a single B-DNA molecule is basically governed by the phosphate groups
which exhibit a double helix structure with right-hand helicity. We model this
 by an infinitely long neutral
hard cylinder oriented in $z$ direction
with additional charged hard spheres whose centers are located
on top of the cylindrical surface. Each  charged  sphere describes a phosphate
group and hence the spheres form
a double helix structure. In detail, the effective cylindrical diameter $D$ is commonly
chosen to be $D=20\AA$  \cite{lebret,hecht,paulsen}. The spheres are
monovalent, i.e.\ 
their charge $q_p<0$ corresponds to one elementary charge $e>0$, $q_p=-e$, and they
have an effective diameter $d_p$. We do not fix $d_p$ but keep it as an additional
(formal) parameter in the range between  $d_p=0.2\AA$ (practically the
point-like charge limit) to $d_p=6\AA$ (to incorporate a  groove geometry 
for the molecule).
Furthermore,  the helical pitch length is
$P=34\AA$; the number of charged spheres  per pitch length (or per helical turn) is 10.
Consequently, successive charges on the same strand are displaced
by an azimuthal angle of $36^\circ$ 
corresponding to a charge spacing of $3.4\AA$ in $z$ direction.
In a plane perpendicular to the $z$ direction, phosphate groups of the two 
different helices are separated by an azimuthal  angle of
$\phi_s=144^\circ$, see Figure~\ref{xyplane}, fixing the  minor 
and the major helical groove
along the DNA molecule. 

We place the discrete charges on the two different
helices such that two of them fall in a common plane perpendicular to the
$z$ axis, see again Figure~\ref{xyplane}. The total line 
charge density along the DNA molecule is then $\lambda=-0.59e/\AA$.

The second DNA molecule is considered to be parallel to the first one
 in our studies. 
 The separation between the two cylinder origins is $R$, 
we also introduce the surface-to-surface separation $h=R-D$. The position of the
two double helices can be described by a relative angle difference $\phi$ between the
two azimuthal angles describing the position of the bottom helix with respect to a fixed
axis in the $xy$ plane. This is illustrated in Figure~\ref{xyplane}. The relative
orientation $\phi$ is the key quantity in describing the angle
 dependence of the forces induced by the 
helical structure. We remark that we only study a situation where the 
discrete phosphates  from  different DNA strands 
possess the same $z$ coordinates for $\phi =0$. Small shifts in the $z$ coordinate
are not expected to change the results significantly. A further
 parameter characterizing the discrete 
location of the phosphate charges along the strands is the azimuthal
angle $\phi_0$ of a phosphate charge with respect to the cylinder separation vector,
see again Figure~\ref{xyplane}. All results are periodic in $\phi_0$ with
a periodicity of $36^\circ$.

In addition to the DNA molecules we describe the  counterions by charged hard spheres
of diameter $d_c$ and charge $q_c$. The counterions are held at  room
temperature
$T=298K$. Their concentration is fixed by the charge of 
the DNA molecules due to the constraint
of global charge neutrality. Also, additional salt ions with charges 
$q_+$ and $q_-$, modelled as charged hard spheres
of diameters $d_+$ and $d_-$,  are incorporated into our model. 
The salt concentration is denoted
by $C_s$. The discrete nature of the solvent, however, is
neglected completely.

The interactions between the mobile ions and phosphate charges 
are described within the framework
of the primitive model as a combination of excluded volume and Coulomb
interactions screened by the dielectric constant $\epsilon$ of the solvent.
The corresponding pair interaction potential between the different 
charged hard spheres is
\begin{equation}
V_{ij} (r) =\cases {\infty &for $ r \leq (d_i+d_j)/2 $\cr
   {{q_i q_j e^2} \over {\epsilon r}} &for $ r > (d_i+d_j)/2 $\cr}.
\label{1cLMH}
\end{equation}
where $r$ is the interparticle separation and $i,j$ are indices denoting
the different particles species. Possible values for $i$ and $j$ are
$c$ (for counterions), $+,-$  (for positively and negatively charged
salt ions), and $p$ (for phosphate groups).
In addition, there is an interaction potential $V^{0}_{i}$ between the DNA hard cylinder and the free ions
$i= c,+,-$ which is of simple excluded volume form such that these ions
cannot penetrate into the cylinder.

Due to the length
of this paper and the large number of quantities, we summarize  most of our
notation in Table~\ref{tab1}.

\section{target quantities}

Our target quantities are equilibrium statistical averages for the local
counter- and salt ion densities  and the effective forces and torques 
exerted onto the bio-molecules. For that purpose we consider a slightly
more general situation with  $N$ parallel DNA molecules contained in a system
of volume $V$. The cylinder centers are  fixed at positions ${\vec R}_i$ ($i=1,...,N$)
in the $xy$-plane. We further assume that there are $N_c$ counterions and
$N_+,N_-$ salt ions in the same system. By this we obtain partial concentrations
$n_c = N_c/V, n_+ = N_+/V, n_- = N_-/V$ of counter and salt ions.

First we define the equilibrium number density profiles $\rho_j(\vec r)$ $(j=c,+,-)$
of the mobile ions in the
presence of the fixed phosphate groups via 
\begin{equation}
\rho_j({\vec r})= \langle \sum_{i=1}^{N_j} \delta ( {\vec r} - \vec
 r_i^{j} )\rangle ,
\label{denc}
\end{equation}
Here $\{ {\vec r}_i^{j}\}$ denote the positions of the $i$th particle of species $j$.
 The canonical
average  $<...>$  over an $\{ {\vec r}_i^{j} \}$-dependent
quantity $\cal A$ is defined via the classical trace
\begin{eqnarray}
\langle {\cal A} \rangle = &&{1\over
 {\cal Z}}\Bigl\{ \prod_{k=1}^{N_c}\int d^3 r_k^c \Bigl\}
\Bigl\{ \prod_{m=1}^{N_+}\int d^3 r_m^+ \Bigl\}
\Bigl\{ \prod_{n=1}^{N_-}\int d^3 r_n^- \Bigl\}
 \nonumber \\
&& \exp
 \lgroup -\beta\sum_{i=c,+,-} [ V^{0}_{i} + \sum_{j=c,p,+,-}  U_{ij} ]
 \rgroup \times {\cal A}
\label{33}
\end{eqnarray}
Here $\beta= 1/k_BT$ is the inverse thermal energy  ($k_B$ denoting
Boltzmann's constant) and 
\begin{eqnarray}
&&U_{ij} =(1-{1 \over 2}\delta_{ij})\sum_{l=1}^{N_i} \sum_{k=1}^{N_j} V_{ij}( \mid {\vec r}_l^i - {\vec r}_k^j \mid ), 
\label{9999}
\end{eqnarray}
is the total  potential energy of the counter- and salt ions  provided the
phosphate groups are at positions $\{ {\vec r}_n^p\}$ ($n=1,...,N_p$). Finally the
prefactor $1/{\cal Z}$ in eq.(\ref{33}) ensures  correct normalization,
 $<1>=1$.
 Note that the  density profiles $\rho_j(\vec r)$ 
also depend parametrically on the
positions $\{ {\vec r}_n^p\}$ of all the fixed phosphate groups ($n=1,...,N_p$).

Now we define 
the total effective force ${\vec {F}}_i$ per pitch length 
acting onto the $i$th DNA molecule $(i=1,...,N)$.
As known from earlier work \cite{LMH,allah2,allah,allahhsg} it contains 
 three different parts 
\begin{equation}
{\vec F}_i = {\vec F}_i^{(1)} + {\vec F}_i^{(2)} + {\vec F}_i^{(3)}. 
\label{neu4}
\end{equation} 
The first term, ${\vec F}_i^{(1)}$, is 
the direct Coulomb force acting onto all phosphate groups
belonging to one helical turn  of the $i$th DNA molecule as exerted from the 
phosphate groups of  all  the other DNA molecules:
\begin{equation} 
{\vec F}_i^{(1)}= -{\sum_k}^{'} \left( {\vec \nabla}_{{\vec r}_k^p}
  \sum_{n=1; n \not= k}^{N_p} V_{pp} \left( \mid {\vec r}_k^p - {\vec
      r}_n^p \mid \right)
\right) 
\label{F1}
\end{equation} 
where the sum $\sum_k'$ only runs over 10 phosphates belonging to one helical turn of
the $i$th DNA molecule.
This term is a trivial sum of direct interactions.

The second term ${\vec F}_i^{(2)}$ involves 
the electric part of the  interaction between the phosphate groups and the
counter- and salt ions. Its statistical definition is
\begin{equation} 
{\vec F}_i^{(2)}= -{\sum_k}^{'} \left( \langle \sum_{i=c,+,-} \sum_{l=1}^{N_i}
 {\vec \nabla}_{{\vec r}_k^p}  V_{pi}( \mid {\vec r}_k^p - {\vec r}_l^i \mid ) \rangle \right)
\label{8}
\end{equation} 
and describes screening of the bare Coulomb interaction (\ref{F1}) by the counter
 and salt ions.

\twocolumn[\hsize\textwidth\columnwidth\hsize\csname
@twocolumnfalse\endcsname

\begin{table}
\caption{List of key variables}
\begin{tabular}{lc}
%\tableline
  $D$ &  DNA diameter  \\ 
  $d_c$ & counterion diameter \\ 
  $d_p$ & phosphate diameter \\ 
  $d_+,d_-$ & salt ion diameters  \\ 
  $ P $     & helical pitch length   \\ 
  $L$ & length of simulation box  \\ 
  $\epsilon$ & dielectric constant of DNA and water  \\ 
  $T$   &temperature \\
  $N_p$ &  number of phosphates in the simulation box  \\ 
  $N_c$ &  number of counterions in the simulation box  \\ 
  $N_s$ &  number of salt ion pairs in the simulation box \\ 
  $C_s$ &  salt concentration  \\ 
  $q_c$ & counterion valency \\ 
  $q_p$ & phosphate valency \\ 
  $q_+,q_-$ & salt ion valencies  \\ 
  $\lambda$ & linear charge density of the DNA molecule \\ 
  $\lambda_B$ & Bjerrum length \\ 
  $\Gamma_{pc}$ & coupling parameter between phosphates and counterions  \\ 
  $F$ & interaction force per pitch length\\ 
  $F_0$ & used unit for force , $F_0=({e \over 4D})^2 $  \\ 
  $M$ & torque acting onto the  DNA molecules  \\ 
  $R$ &  interaxial separation between DNA molecules  \\ 
  $h$ &  surface-to-surface separation between DNA molecules  \\ 
  $\phi$ &  relative orientational angle between two DNA molecules \\ 
  $\phi_0$ &  reference orientational angle for one DNA molecule\\ 
  $F^{(HC)}$ & interaction force  per pitch length within the
  homogeneously charged cylinder model\\
  $\lambda_D$ & Debye screening length  \\ 
  $F^{(YS)}$ & interaction force  per pitch length within the  Yukawa
  segment model \\  
  $r_p^*$ &  effective phosphate radius in the  Yukawa segment model \\ 
  $q_p^*$ &  effective phosphate charge in the  Yukawa segment model \\ 
  $\zeta$ & size correction factor in the  Yukawa segment model  \\ 
  $F^{(KL)}$ & interaction force  per pitch length within
  Kornyshev-Leikin  theory  \\  
  $\theta$ & condensation parameter of counterions \\ 
\end{tabular} 
\label{tab1}
\end{table}
]

Finally, the third term ${\vec F}_i^{(3)}$ describes a
contact (or depletion)  force arising from the hard-sphere part in
$V_{pi}(r)$ and $V^{0}_{i}$ ($i=c,+,-$).  It
 can be expressed as an integral over the molecular surface 
${\cal S}_i$ associated with the excluded volume per one
helical turn of  the $i$th DNA molecule:
\begin{equation}  
{\vec F}_i^{(3)}=-k_BT \int_{{\cal S}_i} d{\vec f} \ \
\left(\sum_{j=c,+,-} \rho_j({\vec r}) \right ) ,
\label{9}
\end{equation} 
where ${\vec f}$ is a surface normal  vector pointing outwards the DNA molecule.
This depletion term is usually neglected in any linear electrostatic
treatment but
becomes actually important for strong Coulomb
coupling $\Gamma_{pc}$ as conveniently defined by \cite{allah,Hartmut,allahhsg}
\begin{equation}
\Gamma_{pc} ={\mid {q_p \over q_c }\mid}{2 \lambda_B \over
  {d_p+d_c}} ,
\label{gamma_mc}
\end{equation}
with the Bjerrum length  $\lambda_B = q_c^2e^2/\epsilon k_B T$.
When $\Gamma_{pc}$ is much larger than one, the Coulomb interaction
 dominates thermal interactions and counterion condensation may occur.
 For DNA molecules this is relevant as
 $d_p+d_c=4-6\AA$ and $\lambda_B=7.14\AA$ for a monovalent counterion in
 water at room temperature, resulting in a coupling parameter 
 $\Gamma_{pc} $ larger than one.

Our final target quantity is  the total torque  per pitch length
acting onto the $i$th DNA molecule. Its component $M_i$ along the $z$-direction
(with unit vector ${\vec e}_z$) can also be decomposed into three parts
\begin{equation}
M_i = M_i^{(1)} + M_i^{(2)} + M_i^{(3)}
\label{M}
\end{equation}
with
\begin{equation}
M_i^{(1)} =  -{\vec e}_z \cdot {\sum_k}^{'} {\vec r}_k^p \times
 \left({\vec \nabla}_{{\vec r}_k^p} \sum_{n=1; n \not= k}^{N_p}
  V_{pp}( \mid {\vec r}_k^p - {\vec r}_n^p \mid) \right) 
\label{M1}
\end{equation}
\begin{equation}
M_i^{(2)} =  -{\vec e}_z \cdot  {\sum_k}^{'} {\vec r}_k^p \times
\left(\langle \sum_{i=c,+,-} \sum_{l=1}^{N_i}
 {\vec \nabla}_{{\vec r}_k^p}
  V_{pi}( \mid {\vec r}_k^p - {\vec r}_l^i \mid) \rangle   \right )
\label{M2}
\end{equation}
and
\begin{equation}
M_i^{(3)} = k_BT {\vec e}_z \cdot \int_{{\cal S}_i} d{\vec f} \times {\vec r} \ \
\left(\sum_{j=c,+,-} \rho_j({\vec r}) \right )
\label{M3}
\end{equation}

\section{Computer Simulation}
\label{simdet}

\begin{figure}
   \epsfxsize=8cm %6cm
   \epsfysize=10cm%7cm
~\hfill\epsfbox{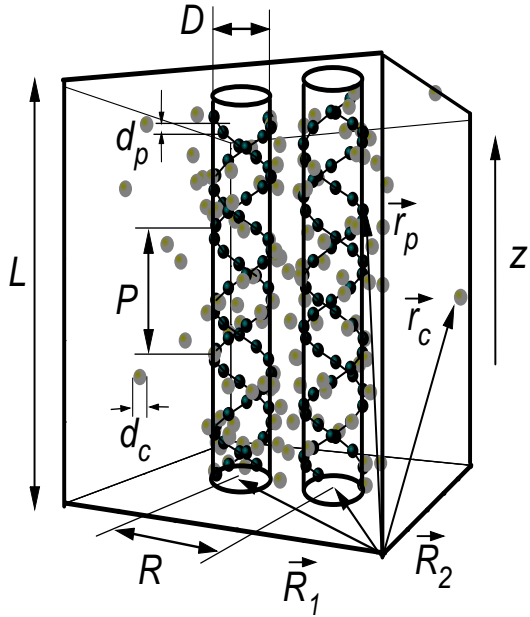}\hfill~
   \caption{Schematic view of the set-up: Two cylindrically shaped DNA
     molecules with a distance $R$ at positions $\vec R_1$ and $\vec R_2$ are placed
     parallel to the $z$-axis inside a  cube of length $L$. The large gray
     spheres are counterions of diameter $d_c$. The black spheres of
     diameter $d_p$, connected by the solid line, are phosphate charges on the 
cylindrical surface 
     of diameter $D$. $P$ is the pitch of DNA\@. Arrays $\vec r_p$ and
     $\vec r_c$ point to positions of phosphates and counterions. 
For sake of clarity, the positions of added salt ions are not
     shown. There
are periodic boundary conditions in all three directions.}
   \label{3dimen}
\end{figure}

Our computer simulation was performed  within a simple  set-up 
which is schematically shown in Figure~\ref{3dimen}.
We consider two parallel DNA molecules in a cubic box of length $L$ with
periodic boundary conditions in all three directions. $L$ is chosen to be three times
the pitch length $P$ such that there are $N_{p}=120$  phosphate charges in the box.
The number of counterions $N_c=120$ in the box is fixed by
charged neutrality while the number of salt ions, $N_s$, is governed by 
its concentration $C_s$. The separation vector between the centers of the
two molecules points along the $x$-direction of the simulation box. The
relative orientation is described according to our notation presented in chapter II,
see again Figure~\ref{xyplane}.

We performed a standard Molecular Dynamic (MD) code with velocity Verlet
 algorithm \cite{allen}.
 System parameters used in our simulations are listed
in Table~\ref{tab2}. The time step  $\triangle{t}$ of the simulation 
was typically chosen to be
$10^{-2}\,\sqrt{m\,{d^{3}_{m}}/e^{2}}$, with $m$ denoting the (fictitious)
mass of the mobile ions,
such that the reflection of
counterions following the  collision with the surface of DNA core cylinder and
phosphates is calculated with high  precision. For every
run the  state of the system was checked during the
simulation time. This was done by monitoring the temperature, average velocity, 
the distribution function of velocities and total potential energy of
the system. On average it took about $10^4$ MD steps to get into
equilibrium. Then during $5\cdot 10^4-5\cdot 10^6$ time steps, we gathered
statistics to perform the canonical averages for calculated quantities.

The  long-ranged nature of the  Coulomb
interaction was numerically treated  via the efficient method proposed by Lekner
\cite{lekner}. A summary of  this method is given in Appendix A. 
In order to save CPU time, the Lekner
forces between pair particles were tabulated in a separate code before
entering into the main MD cycle. The tabulation on a $510 \times 510 \times 510$ grid
with spatial step =$0.1\AA$ was done in the following manner. The first
particle was fixed at the origin (0,0,0) while the second charge was
successively embedded on sites of the generated grid. Then the force
components  acting
onto the first charge were calculated via the Lekner method. A  force data file 
was created which was  used as a common input for all subsequent MD runs.  
 To decrease error
coming from a finite grid length, the forces in the simulations were calculated using
the four-step focusing technique \cite{gilson1}.

\section{Linear screening theory}

Linear screening theory can be used to get explicit 
analytical expressions for the effective interactions
between helical bio-molecules. These kind of theories, however, should
only work for weak Coulomb coupling and thus represent
a further approximation to the primitive model. Depending on the 
form  of the fixed charge pattern characterizing the biomolecules,
one obtains different approximations. 

\subsection{Homogeneously charged cylinder}

The simplest approach is to crudely describe the biomolecule
as a homogeneously charged cylinder. In this case, the effective
interaction force per pitch length between two parallel rods reads \cite{Hansen}
\begin{equation}
{\vec F} \equiv{\vec F}^{(HC)}   =  \frac{2 \lambda^2 P \lambda_D
 K_1(r/\lambda_D)}
{\epsilon {(D /2)}^2  K_1^2( D/(2\lambda_D ))}\frac{\vec r}{r}
\label{ldeb00}
\end{equation}

\twocolumn[\hsize\textwidth\columnwidth\hsize\csname
@twocolumnfalse\endcsname

\begin{table}
\caption{Parameters used for the different simulation runs. The Debye screening length
$\lambda_D$, as defined by Eqn.(\ref{ldeb}), and the Coulomb coupling $\Gamma_{pc}$ are also given.}
\begin{tabular}{lcccccc}
Run& $d_c(\AA)$ & $d_p(\AA)$ & $N_s$ & $C_s(M)$&$\lambda_D(\AA)$&$\Gamma_{pc}$ \\
\tableline
  A &   1  &  0.2  & -    & -     &  9.6 & 12  \\ 
  B &   2  &  2    & -    & -     &  9.6 & 3.6 \\ 
  C &   2  &  6    & -    & -     &  9.6 & 1.8 \\ 
  D &   1  &  0.2  & 15   & 0.025 &  8.6 & 12  \\ 
  E &   1  &  0.2  & 60   & 0.1   &  6.8 & 12  \\ 
  F &   1  &  0.2  & 120  & 0.2   &  5.6 & 12  \\ 
  G &   1  &  0.2  & 440  & 0.73  &  3.3 & 12  \\ 
  H &   1  &  0.2  & 1940 & 3.23  &  1.7 & 12  \\ 
  I &   2  &  2    & 120  & 0.2   &  5.6 & 3.6 \\ 
\end{tabular} 
\label{tab2}
\end{table}
]

Here $r$ is the axis-to-axis separation distance between cylinders, $\lambda_D$ is the Debye-H\"uckel screening length fixed by
\begin{eqnarray}
\lambda_D =\sqrt {\frac{\epsilon k_B T}{4 \pi\gamma
    ( n_c (q_c e)^2 + n_+(q_+ e)^2 + n_-(q_- e)^2 )}}
\label{ldeb}
\end{eqnarray} 
where the factor $\gamma = 1- V_{cyl}/V$ is a correction due to the fact
that the mobile ions cannot penetrate into the cylindric cores 
which excludes a total volume $V_{cyl}$.
Furthermore, $K_1(x)$ is a Bessel function of imaginary argument.
Obviously, the torque is zero for this charge pattern.

\subsection{Yukawa segment model}

It is straightforward to
generalize the traditional Debye-H\"uckel approach to a general
charge pattern  resulting in a Yukawa-segment (YS)
model \cite{stigter,schildkraut,bailey,record1,delrow,soumpasis}.
One phosphate charge interacts with another phosphate charge via an
effective Yukawa potential \cite{vervey}
\begin{equation}
U(r) = \frac{(q_p \zeta)^2 e^2 }{\epsilon r} \exp (-r/\lambda_D )
\label{ldeb99}
\end{equation}
Here, $\zeta $ describes  a size correction
due to the excluded volume of the phosphate groups. This term is assumed to be of
the traditional Derjaguin-Landau-Verwey-Overbeek (DLVO) form 
\begin{eqnarray}
\zeta  =  \exp (r_p^* \lambda_D )/ ( 1 + r_p^* \lambda_D )
\label{qqq}
\end{eqnarray}
where  $r_p^*=(d_p+d_c)/2$ is an effective phosphate radius for the phosphate
counterion interaction.
We remark that nonlinear screening effects and the excluded volume of the cylinder
can also be incorporated by replacing the bare phosphate charge $q_p$ with
 an  effective phosphate charge
$q_p^*$\cite{stigter,bailey,HartmutJCP1994}.

Using the same notation as in chapter III, the total effective force per pitch length
acting onto the $i$th bio-molecule is 
\begin{equation} 
{\vec F}_i \equiv {\vec F}_i^{(YS)}= -{\sum_k}^{'} 
 {\vec \nabla}_{{\vec r}_k^p} \sum_{n=1; n \not= k}^{N_p}
  U ( \mid {\vec r}_k^p - {\vec r}_n^p \mid) 
\label{FYS}
\end{equation} 
within in the Yukawa segment model
where the sum $\sum'$ has the same meaning as in Eqn.(\ref{F1}).
Note that the contact term (\ref{9}) is typically neglected in linear screening theory.
Furthermore, the effective torque per pitch length is
\begin{equation} 
 M_i \equiv M_i^{(YS)}=  -{\vec e}_z \cdot {\sum_k}^{'}
 {\vec r}_k^p \times \left(
 {\vec \nabla}_{{\vec r}_k^p} \sum_{n=1; n \not= k}^{N_p} 
  U( \mid {\vec r}_k^p - {\vec r}_n^p \mid) \right) 
\label{MYS}
\end{equation}
There are also analytical expressions for the equilibrium density profiles
of the mobile ions involving a linear superposition of Yukawa orbitals 
around the phosphate charges \cite{LHM} which, however,  we will not discuss 
 further in the sequel.

\subsection{Kornyshev-Leikin theory}
The linear Debye-H\"uckel screening theory was recently developed further
and modified to account for dielectric discontinuities and counterion adsorption
in the grooves of the DNA molecule by Kornyshev and Leikin (KL)
\cite{kor1,kor2,kor3,kor4,kor5}. An analytical
expression for the effective pair potential $V_{KL}(R,\phi )$ per pitch length
between two parallel rods of separation $R$ with relative orientation $\phi$
was given for separations  larger than $R>D+\lambda_D$.
 Here we only discuss the leading contribution in the special case of
no dielectric discontinuity which reads
\begin{equation}
V_{KL}(R,\phi )  =\frac{8 P  \lambda^2}{\epsilon D^2}
\sum_{n=-\infty}^{\infty}(-1)^n \frac{P_n^2 \cos(n\phi)
  K_0(k_nR) }{k_n^2 (1-\beta_n)^2(K_n^{'}(k_nD/2))^2} 
\label{main}
\end{equation}
and corresponds to the interaction of helices whose strands form
continuously charged helical lines. In Eqn.(\ref{main}),
\begin{eqnarray}
\beta_n= \frac{ng}{k_n}\frac{K_n(k_n
  D/2)I_n^{'}(ngD/2)} {K_n^{'}(k_n D/2) I_n(ngD/2)},
\label{main1}
\end{eqnarray}
\begin{eqnarray} 
 k_n=\sqrt{1/\lambda_D^2 + (ng)^2},\,\,g=\frac{2\pi}{P} ,  
\label{main2}
\end{eqnarray}
 $K_n$ and $I_n$ are modified Bessel
 functions of $n$th order, and $K_n^{'}(x)=dK_n(x)/dx$,
 $I_n^{'}(x)=dI_n(x)/dx$.  

We emphasize that the KL-theory does not assume a priori the double
helical phosphate charge pattern as defined in chapter II\@.
There are rather more possible charge patterns considered 
including a condensation of counterions in the minor and major groove
along the phosphate strands, and on the cylinder as a whole. This involves
four phenomenological parameters as a further input for the KL theory
which makes a direct comparison to the simulation data difficult. In fact,
for the charge pattern given in chapter II, the KL-theory reduces
to the Yukawa-segment model.

In detail, the  charge pattern is characterized by the  form factor $P_n$
\begin{eqnarray}
P_n=&&(1-f_1-f_2-f_3)\theta\delta_{n,0} +\nonumber \\\nonumber
&&f_1\theta+f_2(-1)^n\theta-(1-f_3\theta)\cos(n\phi_s). 
\end{eqnarray}
Here  $\delta_{n,m}$ is the Kronecker's delta function; 
$\theta$ is the first  phenomenological input parameter which
describes the fraction of counterions that are condensed on the whole  cylinder. 
The three numbers $f_i$
 denote the fractions of counterions in the middle of the minor groove
 ($f_1$), in the middle of the major groove ($f_2$), and on the phosphate
 strands ($f_3$) with respect to all condensed counterions. We note that 
the sum in (\ref{main}) rapidly converges, such that  it can  safely be truncated
 for  $\vert n \vert >2$. 
It is straightforward to obtain the  effective  force and torque per
 pitch length between two  molecules 
from (\ref{main}) by taking gradients with respect to $R$ and $\phi$.

\section{Results for point-like charges and no added salt}

In what follows, we consider the set-up of two parallel bio-molecules with
periodic boundary conditions
shown in Figure 2. We projected $\vec F_1$ onto the vector $\vec R$, defining $F =
\vec F_1 \cdot ({\vec R_1} - {\vec R_2})/ \mid
{\vec R_1} - {\vec R_2}\mid $.  Hence a negative sign of $F$ 
implies attraction, and a positive sign  repulsion.
The torque is given for the first DNA molecule, hence $M\equiv M_1$.
We start with the case of no added salt. First, we assume 
the counterion and phosphate diameters to be small, in order to formally
investigate the system with a high coupling parameter  $\Gamma_{pc}
> 10$. 

\subsection{Distribution of the counterions around  the DNA molecules}
\label{sid}

\begin{figure}
   \epsfxsize=8cm
   \epsfysize=10cm %5cm
~\hfill\epsfbox{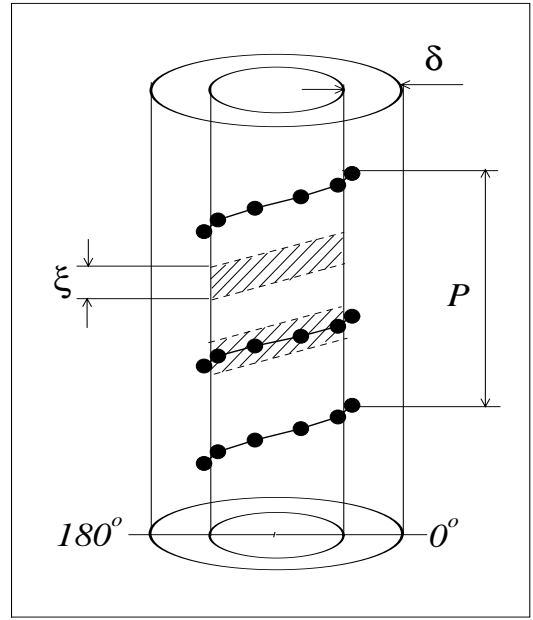}\hfill~
   \caption{A schematic picture to explain the procedure of
     counterion density calculations along one pitch length of a DNA
     molecule. The filled circles connected with solid line are phosphate
groups. The shaded areas correspond to a  path along the major
     groove and along one  phosphate strand. 
The considered volume  has a height $\xi$ and width $\delta$. The
     neighbouring  DNA molecule is assumed to be on the right hand side.}
   \label{cylind}
\end{figure}

We calculated the equilibrium density field (\ref{denc}) of the counterions 
in the vicinity of the DNA molecules by computer simulation. 
In detail, we considered three  different paths to show the counterion density
profile around the first DNA molecule: along a phosphate strand
and along the minor and major groove.
In order
to reduce the statistical error we course-grained this density field further
in a finite volume which is illustrated in Figure~\ref{cylind}. 

\begin{figure}
%   \epsfxsize=9cm
%   \epsfysize=11cm 
%~\hfill\epsfbox{dens_1.ps}\hfill~
%\label{dens-1}
%\end{figure}
%\begin{figure}
   \epsfxsize=8cm
   \epsfysize=10cm %5cm
~\hfill\epsfbox{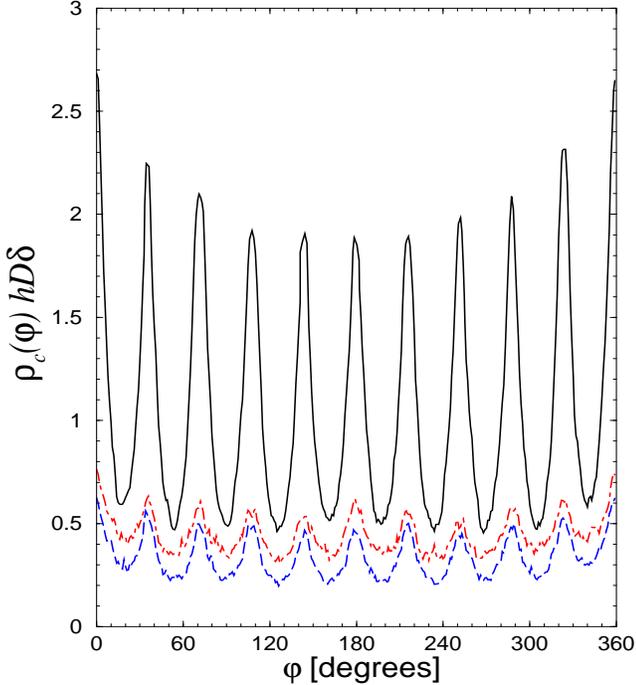}\hfill~
   \caption{Equilibrium counterion density profile 
      $\rho_c(\varphi)$ in units of $1/hD\delta$ versus azimuthal 
     angle $\varphi$ for the parameters of run A,  $\phi=0^{\circ}$ and
a rod separation of  $R=30\AA$.  Solid line:  counterion
     density profile along a phosphate strand (due to symmetry, the
     counterion density profiles on the two phosphate strands are  the
     same).  Dashed line:  counterion density profile
     along the  major  groove. Dot-dashed line:  counterion density profile
     along the minor groove.}
   \label{dens-1}
\end{figure}

This volume is winding around the molecules with a height $\xi$ and width $\delta$.
We choose $\xi=3.4\AA$ and $\delta=2\AA +d_c/2$. In Figure~\ref{dens-1} we plot 
this coarse-grained density field $\rho_c(\varphi)$
 versus the azimuthal angle  angle $\varphi$ 
 from $0^{\circ}$ to $360^{\circ}$ where $\varphi$ is $0^{\circ}$ resp.\  $360^{\circ}$
 in the inner region between the DNA molecules.

Obviously, the counterion density profile has maxima in the 
neighbourhood of the fixed phosphate
charges. Furthermore the concentration of counterions is higher in the minor
than in the major grooves with the $\varphi$-dependence reflecting again the
position of the phosphate charges. Also in the inner region between the two DNA molecules,
there are on average more counterions than in the outside region.

\subsection{Nearly touching configurations}

\begin{figure}
   \epsfxsize=8cm  %6
   \epsfysize=10cm  %7cm
~\hfill\epsfbox{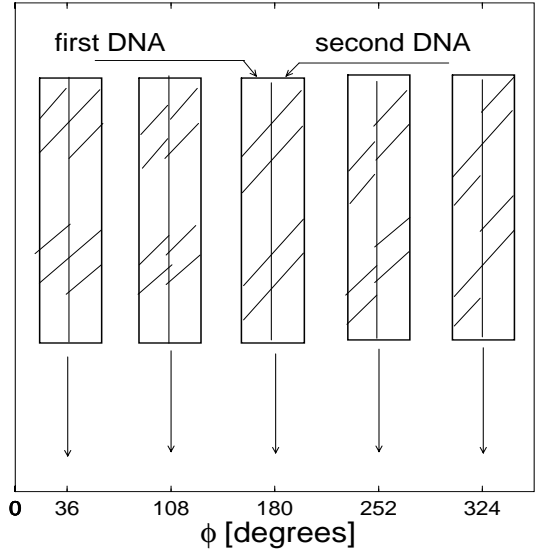}\hfill~
   \caption{Schematic picture of a DNA-DNA configuration for close
     separation distances. The abscissa corresponds to the rotation
     angle of the first DNA molecule. The second DNA molecule is fixed.}
   \label{sxema}
\end{figure}

Let us now consider very small surface-to-surface 
separations between the DNA molecules. In this case one 
expects that the dependence of the forces and torques  on the relative
orientation $\phi$ is most pronounced. For such nearly touching
configurations, however, the discreteness of the phosphate charges,
as embodied in the parameter $\phi_0$, strongly influences the results as well.
The qualitative behaviour of the  $\phi$ dependence can be
understood from Figure~\ref{sxema}. Here two touching DNA molecules
are shown for different relative orientations $\phi$
where the phosphate strands are schematically drawn as continuous lines.
For certain angles $\phi$ which we call touching angles, two neighbouring
phosphate charges hit each other. Possible touching angles are $\phi=36^{\circ},
180^{\circ},324^{\circ}$. 
If $\phi_0$ is chosen to be zero, then
two point charges are  opposing eachother directly. Hence a strong dependence on 
$\phi$ and on $\phi_0$ is expected near touching angles.

\begin{figure}
   \epsfxsize=8cm  %6cm
   \epsfysize=10cm %7cm
~\hfill\epsfbox{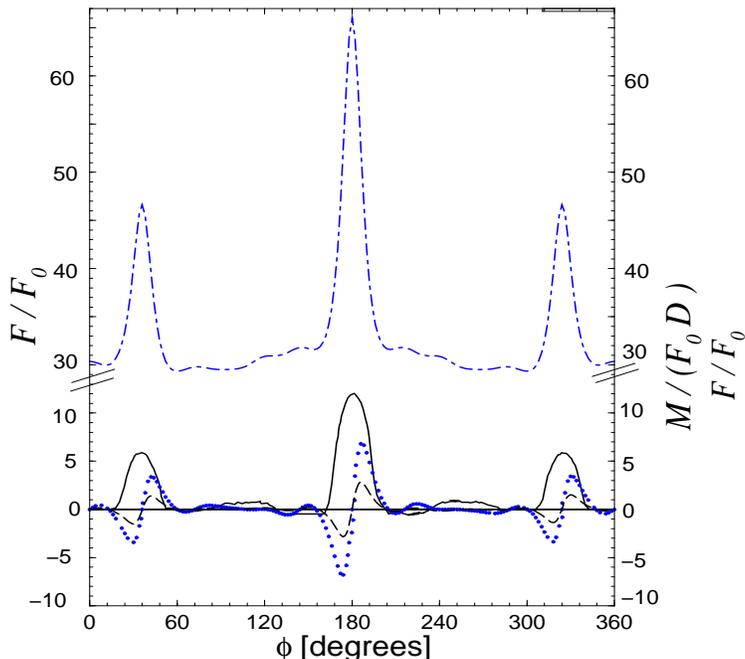}\hfill~
   \caption{Interaction force $F$( left
     $y$-axis)  and  torque $M$ (right $y$-axis)  
     for fixed surface distance $h=2\AA$ versus relative orientation $\phi$
  in degrees. The unit of the force is 
$F_0=(\frac {e}{4D})^2$. The solid (dashed) line is the  simulation result for $F$
($M$) while the dot-dashed (dotted) line
   are data from YS-theory for $F$
($M$). $\phi_0$ is chosen to be zero. The counterion diameter is $d_c=0.8\AA$.}
   \label{fpoint-fi-1}
\end{figure}

Results from  computer simulation
and YS-theory are presented in Figure~\ref{fpoint-fi-1}. The parameters are
from run A (see Table II) but with  $d_c=0.8\AA$. The surface-to-surface
separation is $h=2\AA$.

 For touching angles, the interaction force becomes
strongly repulsive. The strongest repulsion is achieved
for $\phi=180^{\circ}$  since two phosphate strands are meeting simultaneously.
For relative orientations different from a touching angle, the force
becomes smaller and can be both, attractive and repulsive.
YS-theory always predicts a repulsive force. Again there are strong peaks
for touching angles in qualitative agreement with the simulation.
 The actual numbers predicted by YS-theory, however,
are much too large and off by a factor of 6-7 around touching angles.

The torque shows an even richer structure as a function of $\phi$. Near
a touching angle it  exhibits three zeroes corresponding to an unstable minimum
exactly at the touching angle and two stable minima near the touching angles.
The YS-theory shows 2 times larger values for the torque as compared to
the simulation data.

A qualitatively different  force-angle behavior is observed for a larger
counterion diameter. Results for $d_c=1\AA$ are shown in  Figure~\ref{fpoint-fi-2}.

Here at touching angles, the interaction force is attractive. The physical
reason for that are the  contact forces as given by Eqn.(\ref{9}). Caused
by the larger counterion diameter, counterions are stronger depleted 
in the zone  between the DNA molecules. 
The torque has qualitatively the same behaviour as before.

We emphasize that the results do also depend strongly on $\phi_0$.
 For $\phi_0=18^{\circ}$, for instance,  the force
$F$ practically vanishes for any relative orientation $\phi$ as
compared to the same data for $\phi_0=0^{\circ}$.

\subsection{Distance-resolved forces}

We now discuss in more detail the distance-resolved effective forces.
For the parameters of run A, simulation results for $F$ 
are presented in Figure~\ref{fpoint-r}. 

\begin{figure}
   \epsfxsize=8cm  %6cm
   \epsfysize=10cm %7cm
~\hfill\epsfbox{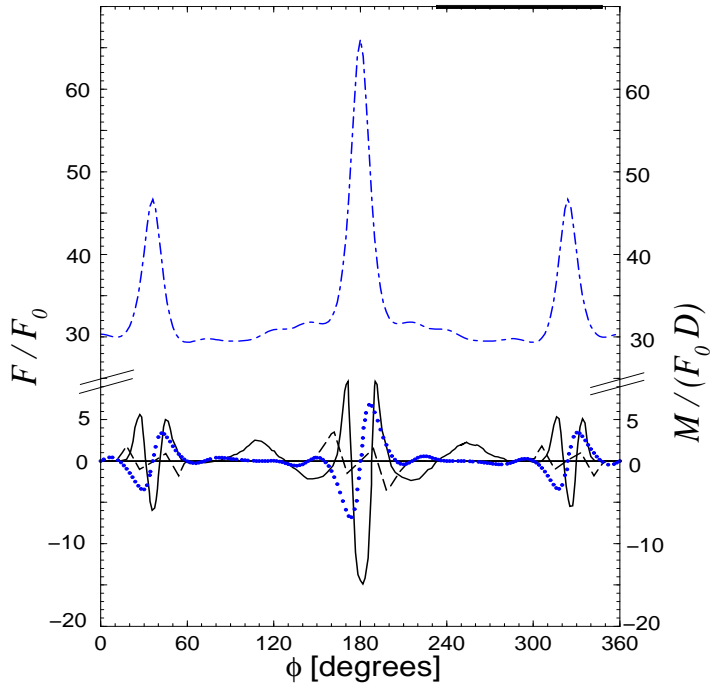}\hfill~
   \caption{Same as Figure~\ref{fpoint-fi-1} but now for $d_c=1\AA$.}
   \label{fpoint-fi-2}
\end{figure}

\begin{figure}
   \epsfxsize=7.5cm
   \epsfysize=9cm %5cm
~\hfill\epsfbox{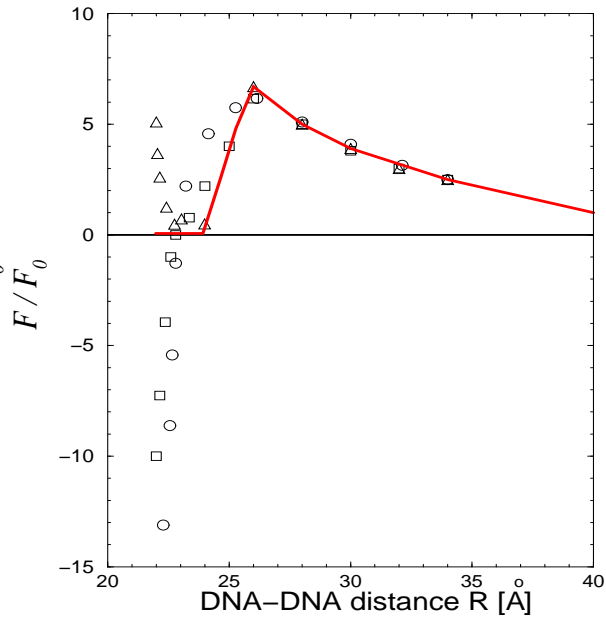}\hfill~
   \caption{Effective interaction force $F$ acting onto a DNA
     pair versus the center-to-center distance $R$.
 The solid line is for  $\phi_0 = 18^{\circ}$. In this case there is no
significant $\phi$-dependence.  The meaning of the symbols, that
 correspond to $\phi_0 = 0$, is :  
circles-  $\phi=180^{\circ}$, squares-  $\phi=36^{\circ}$, 
triangles-  $\phi=45^{\circ}$.}
   \label{fpoint-r}
\end{figure}

For $\phi_0=0$, the force depends on the relative orientation $\phi$
up to a surface-to-surface separation $h\approx 6\AA$ in accordance with 
Figure~\ref{fpoint-fi-2}.
 On the other
hand, for $\phi_0=18^\circ$, there is no $\phi$ dependence at all
for any separation. This supports the conclusion of previous works
\cite{hochberg,conrad}, that the effect of discreteness of the DNA phosphate
charges on the counterion concentration profile is  small in general and
dwindles a few Angstroms from the DNA surface. In fact, for $h> 6\AA$,
there is neither a $\phi$ nor a $\phi_0$ dependence of the  force,
and the total force is repulsive.

\begin{figure}
   \epsfxsize=8cm %6cm
   \epsfysize=10cm %7cm
~\hfill\epsfbox{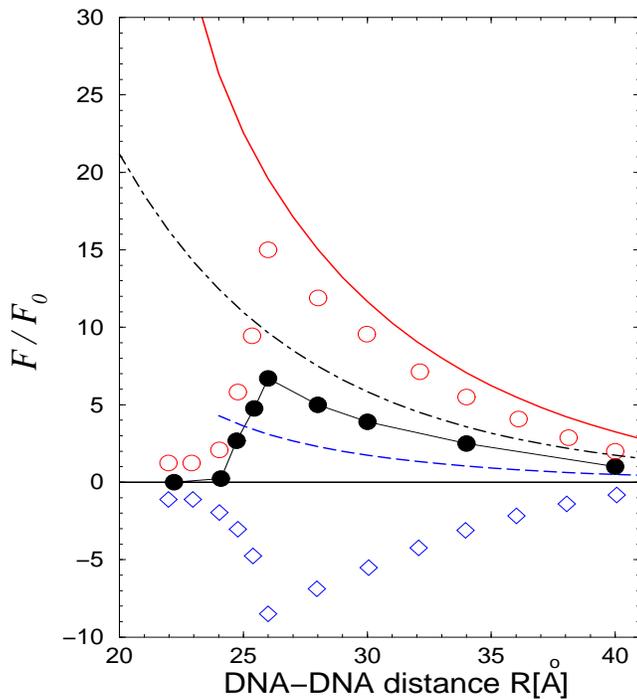}\hfill~
   \caption{Theoretical  and simulation results for
     interaction force $F$ versus separation distance $R$. The unit of the force is 
$F_0=(\frac {e}{4D})^2$.
The parameters are from
run A and $\phi_0 = 18^\circ$.
\\ Symbols: $\bullet$ - simulation data for all DNA rotation
  angles, $\circ$ - the entropic part $ \vec F^{(3)}$, $\diamond$ - the
  pure electrostatic part $ \left( \vec F^{(1)} +\vec
    F^{(2)} \right)$. Solid line: YS theory.
Dot-dashed line: homogeneously charged cylinder model.  
Dashed line: the predictions of KL theory 
 with $f_1=0.1, f_2=0.1, f_3=0.7, \theta=0.71$.}
\label{fmdtheor}
\end{figure}

Furthermore we compare our simulation results with the prediction of
linear screening theories in Figure~\ref{fmdtheor}. First of all, our 
simulation data for the total force (solid circles)
are decomposed into the electrostatic part 

 $ F^{(1)}+F^{(2)}$ (diamonds) and the contact (or depletion) part
$ F^{(3)}$ (open circles). While the latter is strongly repulsive, the
electrostatic part is attractive such that the net force is repulsive.
Linear screening theories aim to describe the pure electrostatic force only.

Results for linear screening theories on different levels are also collected in 
 Figure~\ref{fmdtheor}. If one compares with the {\it total}\/ force,
the prediction
obtained by a homogeneously charged cylinder is repulsive and off by a
 factor of roughly  
1.5. A simulation with a homogeneously charged rod yields perfect
 agreement with linear screening theory since the Coulomb coupling is
 strongly reduced as the rod charges are now in the inner part of the
 cylinder.
  The Yukawa-segment theory is repulsive and off by a factor of 3. 
It is understandable that the YS model leads to a stronger repulsion than the
charged cylinder model as the separation of the phosphate charges in the inner
region between the DNA molecules is shorter than the rod center separation.

\begin{figure}
   \epsfxsize=8cm
   \epsfysize=10cm %5cm
~\hfill\epsfbox{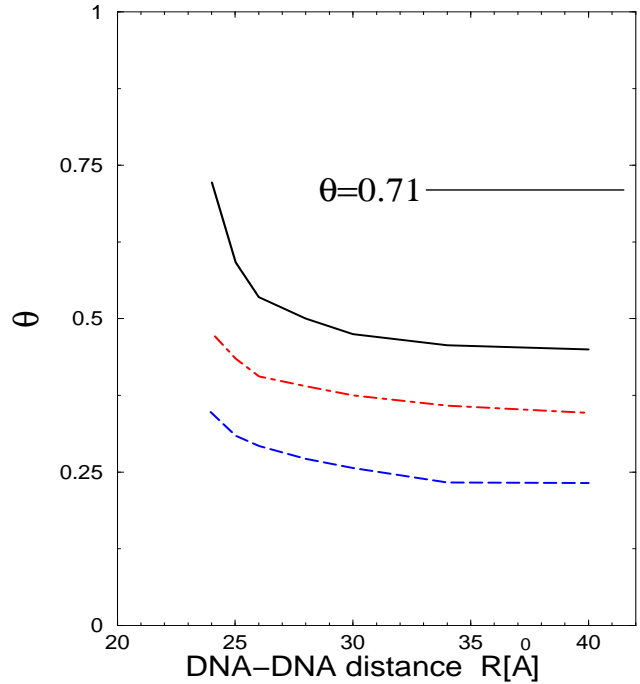}\hfill~
   \caption{The condensation parameter $\theta$ versus separation
     distance $R$. From top to bottom:
solid line- run A ($d_c=1\AA$, $d_p=0.2\AA$),
     dot-dashed line-run B ($d_c=2\AA$, $d_p=2\AA$), dashed line- run C
     ($d_c=2\AA$, $d_p=6\AA$). The horizontal line at $\theta=0.71$
indicates the saturation value at large distances for  a larger 
     $\delta=l_B=7.1\AA$. This saturation value is the same for run A,B, and C.}
   \label{theta}
\end{figure}

The Kornyshev-Leikin
theory requires  four counterion condensation  
fractions $\theta$, $f_1$, $f_2$, $f_3$ as an input.
We have tried to determine these parameters from our simulation in order to get
a direct comparison without any fitting procedure. In order to do so, 
we introduce a small shell around the cylinder of width $\delta$ and
determine $\theta$ as the fraction of counterions which are condensed
onto the DNA within this shell. The actual value for $\delta$ is somewhat
arbitrary,  we first took  a microscopic shell of width
$\delta=2.5\AA$ as well as $\delta=\lambda_B=7.1\AA$.
Data for $\theta$ versus the rod separation are included in Figure~\ref{theta}
for three different combinations of counterion and phosphate diameters. It
becomes evident that the fraction $\theta$  of condensed counterions decreases with  the
rod distance but saturates at large separations.  $\theta$ also depends on the
size of the counterions and phosphate charges.
If the width of the shell $\delta$ is enhanced towards $\delta=\lambda_B=7.1\AA$,
$\theta$ increases again. 
On the other hand, $\theta$ is independent of the relative orientation $\phi$. 
The actual data are  consistent with  Manning's condensation
parameter \cite{manning,manning1} $\theta_0 = \lambda/|q_c|\lambda_B=0.71$
particularly if the width $\delta$ is taken as one Bjerrum length. 
Our data are also in semiquantitative  accordance with
other computer simulations  \cite{jaya1} and 
 nuclear magnetic resonance (NMR)
 experiments which show that the condensed counterion fractions are in
the range of 0.65 to 0.85 \cite{bleam} or  0.53 to 0.57  \cite{padman,mills}.

According to our  results for the counterion density
distribution (see Figure~\ref{dens-1}) we fix the minor and major groove fractions to
$f_1=0.1,f_2=0.1$, and the strand fraction to $f_3=0.7$. Thus, $(1-f_1-f_2-f_3)=0.1$
is the fraction of the condensed counterions which is distributed
neither on the phosphates strands nor on the minor and major grooves. 
The force in KL theory depends sensitively on $\theta$ but is rather insensitive
with respect to $f_1$, $f_2$, $f_3$, and $\phi$.  If the Bjerrum length is taken 
as a width for the condensed counterions, $\theta=0.71$, then the KL
theory underestimates the total force. If, on the other hand, a reduced value
of $\theta =0.545$ is heuristically assumed, then the KL theory reproduces 
the total force quite well.

A serious problem of the comparison with linear screening theories is that
the contact term is not incorporated in any theory apart from recent modifications
\cite{gilson,allah2}. In fact, one should better compare the pure electrostatic
part which is attractive in the simulation. Consequently, none of the linear screening
theories is capable to describe the force well. This is due to the neglection
of correlations and fluctuations in linear screening theories. From a more
pragmatic
point of view, however, one may state that a suitable charge renormalization
leads to quantitative  agreement with the {\it total\/} force. In fact, all three
theories yield perfect agreement if the phosphate charges resp.\ the condensation
parameter $\theta$ is taken as a fit parameter. For instance,
the YS-model yields perfect agreement with the simulation 
for distances larger than $26\AA$
if in Eqn.(\ref{ldeb99}) a renormalized phosphate charge
 $q_p^*=-0.6e$ is taken replacing the bare charge $q_p$.
But this is still unsatisfactory from a more principal point a view.

\section{Results for the grooved model}

\begin{figure}
   \epsfxsize=8cm
   \epsfysize=10cm %5cm
~\hfill\epsfbox{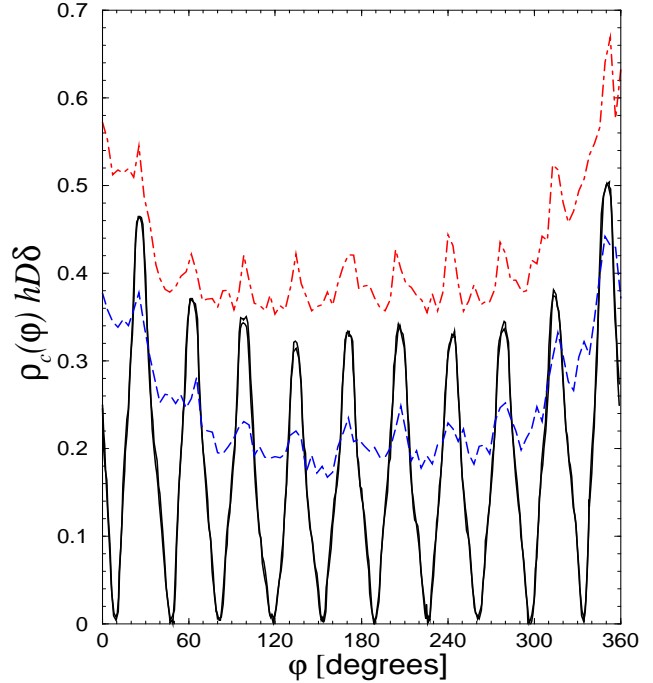}\hfill~
   \caption{Same as Figure~\ref{dens-1} but now for run C and 
 $\phi=45^{\circ}$,  $\delta=3\AA$.}
   \label{densi}
\end{figure}

The groove  structure of DNA is expected to be of  increasing significance as one
approaches its surface \cite{montoro}. We incorporate this in our model by
increasing the phosphate diameter towards  $d_p= 2\AA$ (run B) and 
$d_p= 6\AA$ (run C). Results for  the condensation
parameter $\theta$ are shown in Figure~\ref{theta}.  
$\theta$ is decreasing with increasing $d_p$   since   the coupling
parameter $\Gamma_{pc}$ is decreasing which weakens counterion  binding
to the phosphate groups. Also the qualitative shape of the counterion density profiles
depends sensitively on the groove nature as can be deduced from Figure~\ref{densi}
as compared to Figure~\ref{dens-1}. The counterion density
along the phosphate strands now exhibits  minima at the phosphate charge positions
while it was maximal there in Figure~\ref{dens-1}. Furthermore, 
the counterion density in the minor grooves is now
higher than along the strands due to the geometrical constraints for the counterion
positions which is similar to results of Ref.\ \cite{conrad}.
 In fact, recent X-ray
diffraction \cite{shui1,shui2,mcf} and  NMR spectroscopy \cite{hud1,hud2}
experiments, as well as  molecular mechanics \cite{young1,young2}  and Monte Carlo
 simulations  \cite{klein} suggest that
monovalent cations selectively partition into the minor groove. This effect
is present also  in our simple model and can thus already  be understood from 
electrostatics and thermostatics.

\begin{figure}
   \epsfxsize=8cm  %6cm
   \epsfysize=10cm %7cm
~\hfill\epsfbox{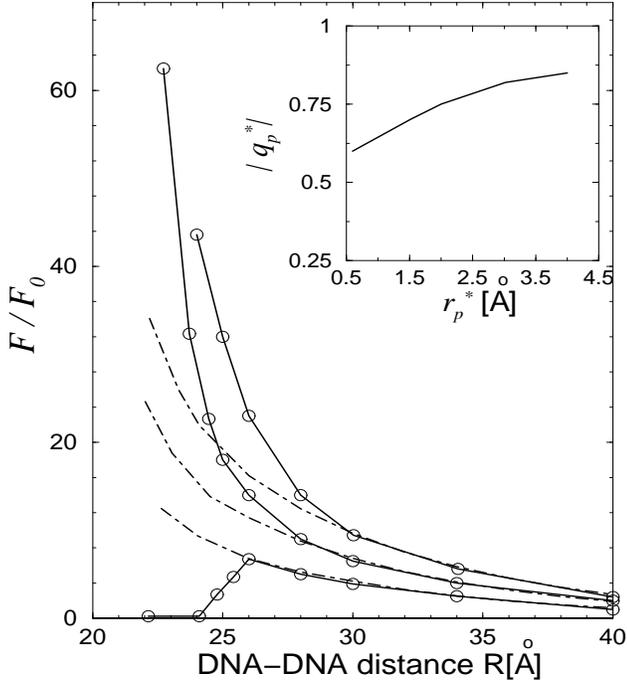}\hfill~
   \caption{Interaction force $F$ versus separation distance $R$.
\\ The open circles  are  simulation data for all relative
orientations $\phi$  with $\phi_0 = 18^{\circ}$. From bottom to top:
$d_c=1\AA$, $d_p=0.2\AA$ (run A);  $d_c=2\AA$, $d_p=2\AA$ (run B); 
  $d_c=2\AA$, $d_p=6\AA$ (run C).\\
  The dashed lines are fits by the YS model.  From bottom to top: 
 fit for the parameters of run A with $q_p^*=-0.6e$; 
 fit for the parameters of run B with $q_p^*=-0.75e$; 
 fit for the parameters of run C with $q_p^*=-0.85e$. The inset is the 
variation of the renormalized phosphate charge $q_p^*$  versus effective
phosphate  radius ${r_p}^*$.}
   \label{fnepoint}
\end{figure}

An increasing phosphate and counterion size increases the 
effective forces which is shown in Figure~\ref{fnepoint}.
Here, as $\phi_0$ was chosen to be $18^\circ$, there is no 
notable dependence on the relative orientation $\phi$.
A similar behavior was observed in a hexagonally
ordered DNA system via Monte Carlo  calculations \cite{lyubartsev}.
This is understandable as counterion screening is becoming less effective.
We have tried to fit the simulation data using  a renormalized charge in the YS theory.
A good fit was obtained for large separations while there are increasing deviations
at shorter distances. This is different from our results for small ion sizes
also shown in Figure~\ref{fnepoint} where the fit was valid 
over the whole range of separations. The adjustable parameter $q_p^*$ 
is shown versus the effective phosphate radius $r_p^*$ of the YS model
in the inset of Figure~\ref{fnepoint}. It is increasing  with increasing $r_p^*$
in qualitative agreement with charge renormalization 
models \cite{AlexanderPincusJCP1984}.

We also note that the physical nature of the electrostatic part of
the interaction force undergoes a transformation upon decreasing the coupling parameter
$\Gamma_{pc}$. For strong
coupling, $\Gamma_{pc}=12$ (run A), the electrostatic part 
$F^{(1)}+F^{(2)}$ is attractive (see Figure ~\ref{fmdtheor}). For
moderate coupling,  $\Gamma_{pc}=3.6$ (run B), it is nearly zero for
all distances. Finally, for weak coupling,$\Gamma_{pc}=1.8$ (run C) the
electrostatic part is elsewhere repulsive. 
The entropic part $F^{(3)}$ for these three runs is always repulsive and
does not undergo a significant change.       

\section{Results for added salt}

Interactions involving nucleic acids are strongly dependent on salt
concentration. Indeed, the strength of binding constants can
change by orders of magnitude with only small changes in ionic
strength \cite{manning2,record6}. Our simulations show a similar strong salt impact on
the interaction force.

When salt ions are added, there is a competition between two effects. 
The first one is the increasing of the direct repulsion between
molecules as a consequence of {\it delocalizing} the  adsorbed counterions. The
second stems from the osmotic pressure of added salt that pushes the salt
ions to occupy the inner molecular region and to {\it screen} the DNA-DNA
repulsion. As we shall show below, these two effects result in a 
novel non-monotonic behaviour of the force as a function of salt concentration.

\begin{figure}
   \epsfxsize=8cm %6cm
   \epsfysize=10cm %7cm
~\hfill\epsfbox{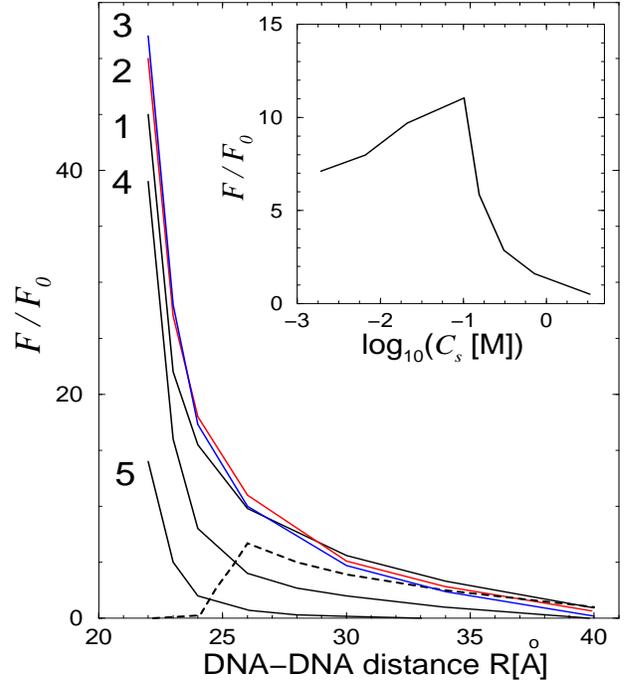}\hfill~
   \caption{Interaction force $F$ acting onto a DNA
     pair versus distance for $\phi=0^\circ$ and $\phi_0 = 18^\circ$.
The unit of the force is 
$F_0=(\frac {e}{4D})^2$.
 The solid lines are for increasing salt concentration: 1- run D, 2 - run E, 
3 - run F, 4 - run G, 5 - run H. 
Dashed line: reference data without salt from run A.
 The inset shows the force versus salt concentration at
fixed separation $R=26\AA$.}
   \label{fsalt}
\end{figure}

 Simulation results for $F$ versus distance for increasing 
 salt concentration are presented in Figure~\ref{fsalt}. In our
simulations,  counter and equally charged salt ions are
 indistinguishable.
 We take $d_+=d_-=d_c, \vert q_+\vert  = \vert q_- \vert = e$.
It can be concluded from Figure~\ref{fsalt} that
 even a small amount of salt ions (line 1, run D,
$C_s=0.025M$) significantly enhances the DNA-DNA repulsion (compare with  the
dashed line corresponding to run A, $C_s=0M$). Upon increasing the salt
concentration, at large separations, $h>10\AA$, the screening is
increased in accordance with the linear theory. However, at intermediate and
nearly touching separations, a non-monotonic behaviour as a function
of salt concentration is observed as illustrated in  the inset of  Figure~\ref{fsalt}. 
In the inset, the maximum of  $F$ occurs for  $C_s=0.2M$. 
The physical reason for that is that 
 added  salt ions first delocalize bound  counterions which
leads to a stronger repulsion. Upon further
increasing the salt concentration, the electrostatic screening is enhanced
again and the force gets less repulsive. In order to support this
picture we show typical microion configurations and investigate also 
the fraction $\theta$ of condensed counterions as a function of salt concentration.

\begin{figure}
   \epsfxsize=8cm %6cm
   \epsfysize=10cm %7cm
~\hfill\epsfbox{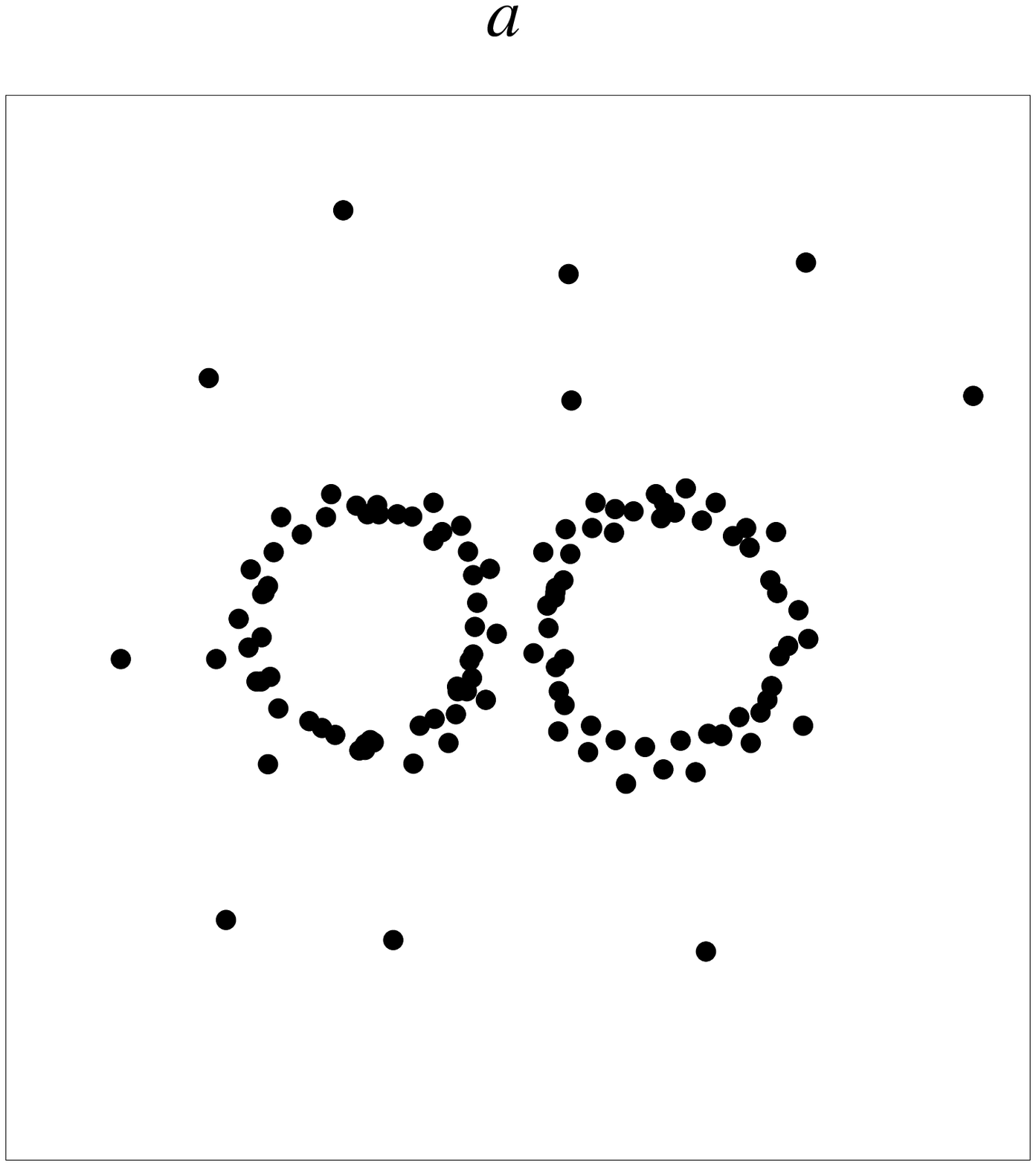}\hfill~
   \epsfxsize=8cm %6cm
   \epsfysize=10cm %7cm
~\hfill\epsfbox{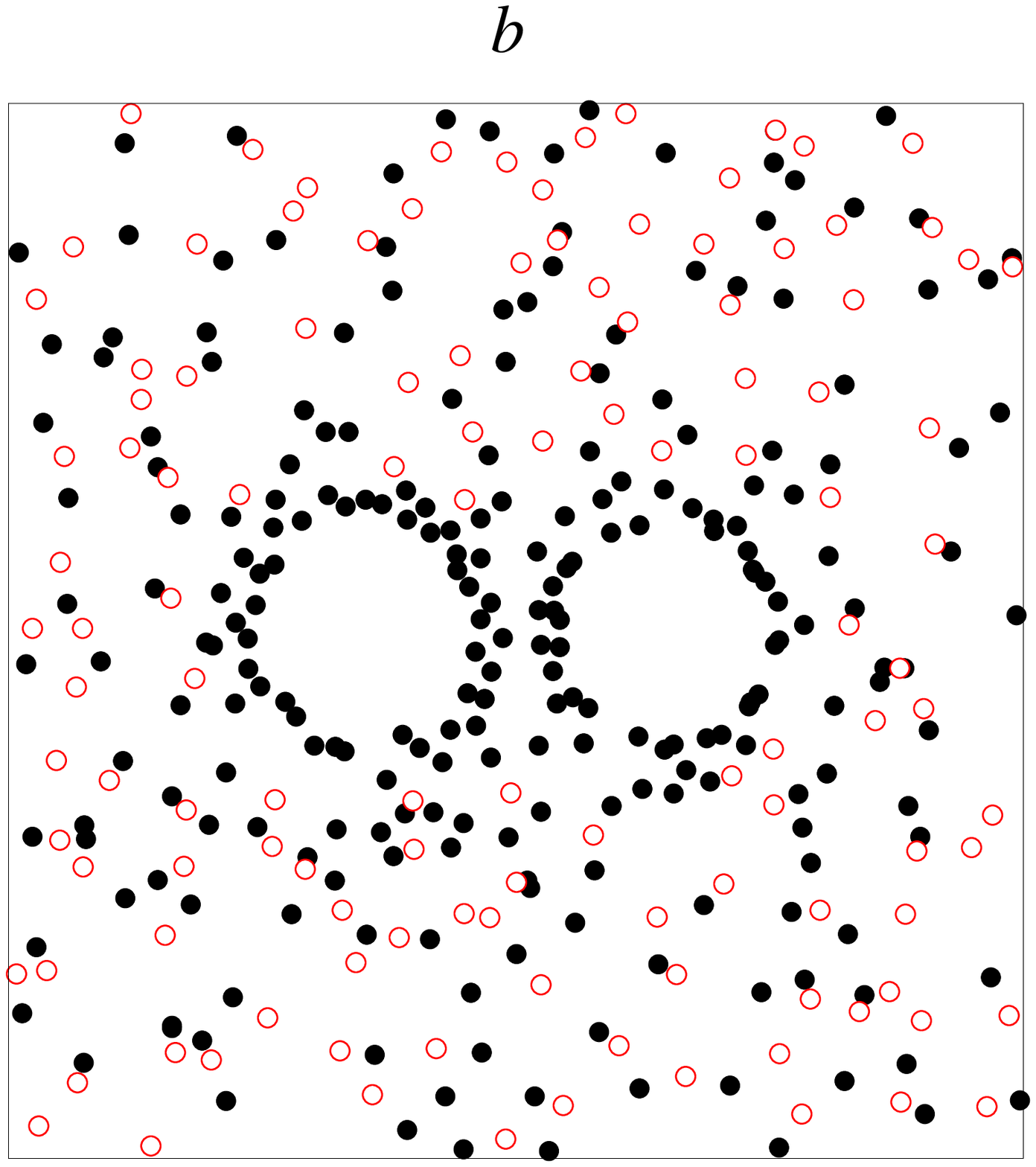}\hfill~
\newpage
%   \label{snap-1}
%\end{figure}
%\begin{figure}
   \epsfxsize=8cm %6cm
   \epsfysize=10cm %7cm
~\hfill\epsfbox{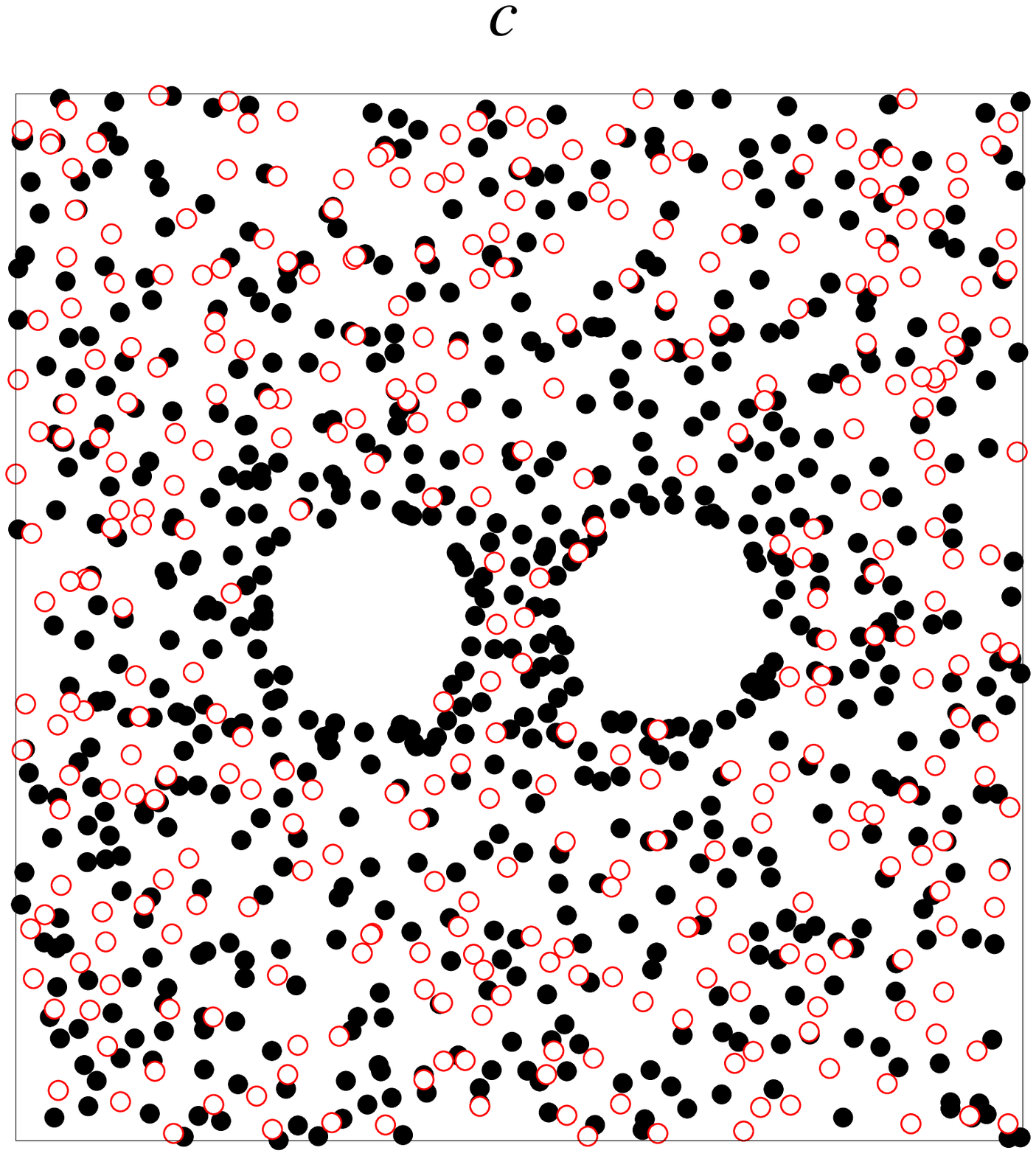}\hfill~
   \caption{Two-dimensional microion snapshots projected to  a plane
     perpendicular to the helices for $\phi=0^\circ$,
     $\phi_o=18^\circ$, $R=30\AA$.  The filled circles are the positions of
     the counterions and positive salt ions, the open circles are the
     positions for the negative salt ions (coions). $a$ - run A, 
$b$ - run F, $c$ -run G.}
   \label{snap-2}
\end{figure}

Simulation snapshots are given in Figure~\ref{snap-2},
 where the positions of the mobile ions are projected onto the $xy$-plane. 

A comparison of the salt-free case (Figure~\ref{snap-2}a) with that of moderate
salt concentration ($C_s=0.2M$, Figure~\ref{snap-2}b) reveals that the total
number of adsorbed counterions decreases with increasing $C_s$.
Furthermore, for $C_s=0.2M$ (Figure~\ref{snap-2}b), there are no coions in the inner
DNA-DNA region. Thus salt ions do not participate in 
screening. Consequently, the DNA-DNA interaction, due to delocalization
of counterions, will be enhanced. Contrary to that, for $C_s=0.73M$,
(Figure~\ref{snap-2}c) the salt co- and counterions enter into the inner DNA-DNA
region and effectively screen the interaction force.

\begin{figure}
   \epsfxsize=8cm
   \epsfysize=10cm %5cm
~\hfill\epsfbox{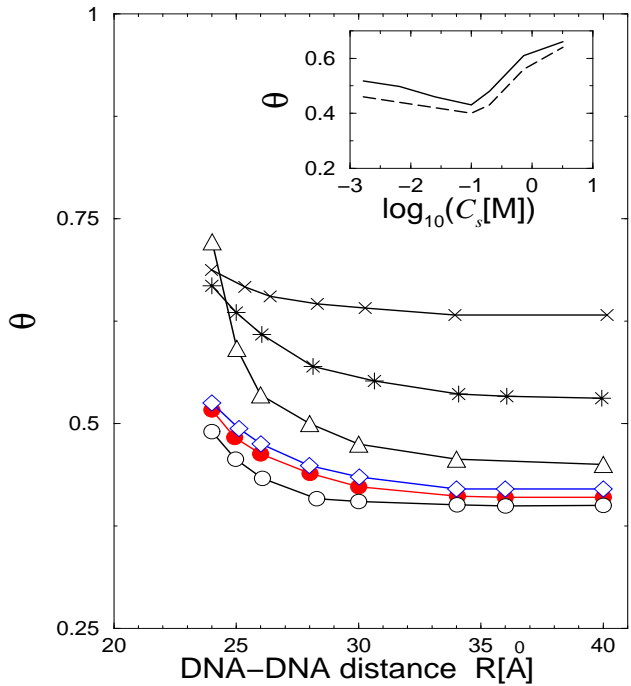}\hfill~
   \caption{Same as Figure~\ref{theta}, but now with added salt.
Symbols:  $\triangle$ - run A, $\bullet$ - run D, $\circ$ - run E,
$\diamond$ - run F, $\ast$ - run G, $\times$ - run H. The inset shows
$\theta$ for fixed distance as a function of  salt concentration:
solid line- for $R=26\AA$; dashed line- for $R=30\AA$.} 
\label{thetasalt}
\end{figure}

Further information is gained from the 
fraction $\theta$ of condensed counterions which is plotted as a function of $R$
 for different
salt concentrations $C_s$   in
Figure~\ref{thetasalt}. We define $\theta$ as the ratio of
condensed counterions coming from the molecules
with respect to the total number of counterions stemming from the molecules.
As $C_s$ increases, the saturation
of $\theta$  occurs at smaller distances. 
In the inset of Figure~\ref{thetasalt} a non-monotonic behaviour of $\theta$
as a function of the added salt concentration is visible which again
is a clear signature of the scenario  discussed above. 
The  increase of $\theta$ above  a certain threshold of salt concentration
is mainly due to a counterion accumulation  outside the grooves.
 A similar trend  was  predicted
by Poisson-Boltzmann \cite{gueron} and Monte Carlo \cite{lebret,murthy} calculations
in  different models.

\begin{figure}
   \epsfxsize=8cm %6cm
   \epsfysize=10cm %7cm
~\hfill\epsfbox{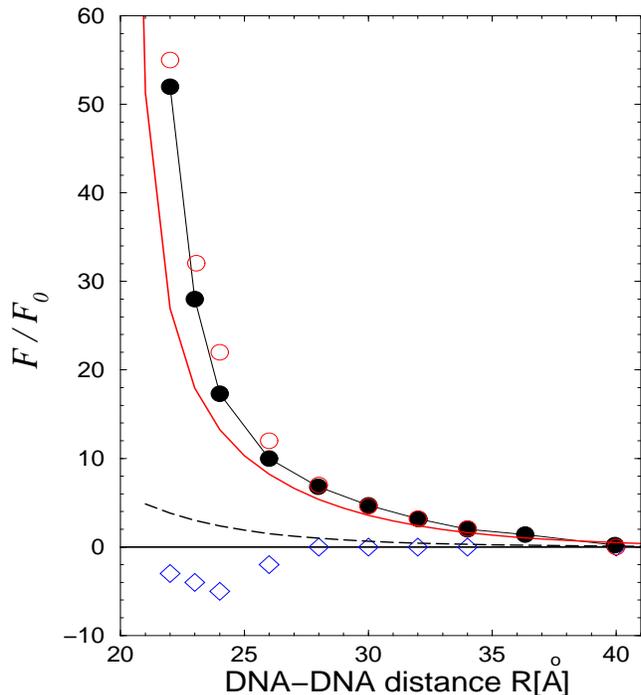}\hfill~
   \caption{Same as Figure~\ref{fmdtheor} but now for run F and 
$\phi = 0^\circ$, $\phi_0 = 18^\circ$. The KL theory was adjusted to
 $f_1=0.1, f_2=0.1, f_3=0.7, \theta=0.71$. The results for KL theory
 and homogeneously charged cylinder models coincide exactly.} 
   \label{fsalt120}
\end{figure}

More details of the forces and the comparison to linear screening theories
are shown in Figures~\ref{fsalt120},~\ref{znorm} and~\ref{f22salt120}.
For run F,  the different parts of the total force are presented in
 Figure~\ref{fsalt120}. As compared to the salt-free case
 (Figure~\ref{fmdtheor}) 
the pure electrostatic part is again attractive  but much smaller,
while the depletion 
part is repulsive and dominates the total force. All three linear
models, homogeneously charged cylinder model, YS, and KL theory, underestimate 
the force. Note that the KL-theory with 
a $\theta$ parameter corresponding to a width $\delta$ of one Bjerrum
length and the homogeneously charged cylinder model give the same results.
 Again with a suitable scaling of the prefactor 
by introducing a renormalized phosphate charge $q_p^*$ resp.\ by fitting 
the condensed fraction $\theta$, one can achieve good agreement with the simulation
data for distances larger than $24\AA$. The fitting parameter $q_p^*$ 
used for the YS-model is $-1.1e$, while the optimal
condensed fraction $\theta$ for the KL-theory is $0.2$. 
The optimal  renormalized phosphate charge $q_p^*$ is
shown versus salt concentration in Figure~\ref{znorm}. Note that the usual DLVO
size correction factor $\zeta$ 
is already incorporated in the interaction, so what one sees are
actual deviations from DLVO theory. The renormalized charge $q_p^*$ increases
with increasing $C_s$ which is consistent with the works of  Delrow {\it et al}
\cite{delrow} and Stigter \cite{stigter}. If one simulates the force
within the homogeneously charged rod model, one finds good agreement
with our simulation data for large separations. Consequently, the
details of the charge pattern do not matter for large salt
concentrations.

We also note that our simulations give no notable
dependence of the force on the relative orientation $\phi$  for $h>6\AA$. 
Only for small separations, $h<6\AA$
there is  a slight dependence in agreement with  Ref.\   \cite{hochberg}.

Finally we show the influence of the ion and phosphate size on the effective force
(for the parameters of run I) in Figure~\ref{f22salt120}. The electrostatic part
of the force is now repulsive but the total force is still dominated
by the depletion part. As far as the comparison to linear screening theories 
is concerned, one may draw similar conclusions as for  Figure~\ref{fsalt120}.
The fitting parameter $q_p^*$ needed to describe the long-distance behaviour within the
YS model does not depend sensitively on the phosphate and ion sizes.
With a suitable scaling of the prefactor one can achieve good agreement with the simulation
data for distance larger than $26\AA$. The fitting parameter $q_p^*$ 
used for the YS-model is $-1.1e$, while the optimal
condensed fraction $\theta$ for the KL-theory is $0.19$. Here again,
simulations of the homogeneously charged cylinder model are in good
agreement with our results obtained for a  double stranded DNA molecule.

\section{Comments and conclusions}

In conclusion, we have calculated the interaction between two parallel
B-DNA  molecules within a ``primitive" model. In particular, 
we focussed on the distance- and orientation-resolved
effective forces and torques as a function of salt concentration. Our main
conclusions are as follows:

\begin{figure}
   \epsfxsize=7cm %6cm
   \epsfysize=8cm %7cm
~\hfill\epsfbox{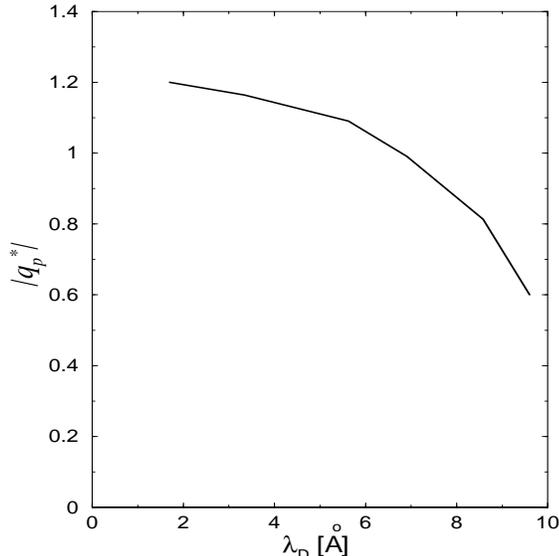}\hfill~
   \caption{Fitted renormalized phosphate charge $q_p^*$
 in the YS
     model,  versus Debye screening length $\lambda_D$ for runs D-H.} 
   \label{znorm}
\end{figure}

\begin{figure}
   \epsfxsize=7cm %6cm
   \epsfysize=8cm %7cm
~\hfill\epsfbox{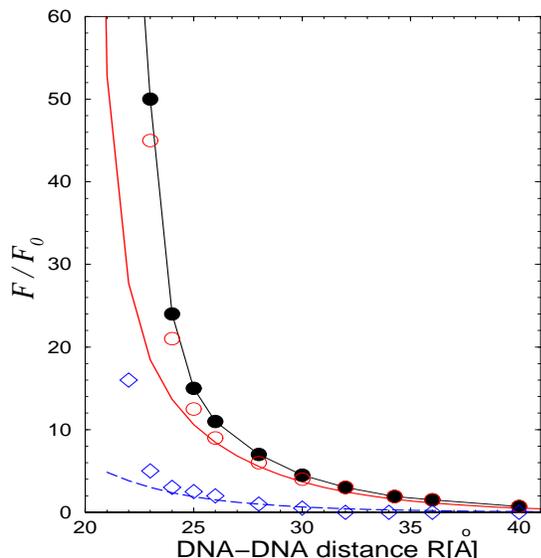}\hfill~
   \caption{Same as Figure~\ref{fmdtheor} but now for run I and 
$\phi = 0^\circ$, $\phi_0 = 18^\circ$. 
The KL theory was adjusted to
 $f_1=0.1, f_2=0.1, f_3=0.7, \theta=0.71$. Note that the KL and
 homogeneously charged cylinder models produce the same curves.} 
   \label{f22salt120}
\end{figure}

First, the interaction force for larger separations is repulsive and dominated
by microion  depletion. The orientational dependence induced by the internal helical
charge pattern is short ranged decaying within a typical surface-to-surface separation
of $6\AA$. For shorter separations there is a significant dependence on the
relative orientation $\phi$ and on the discreteness 
of the charge distribution along the strands.
As a function of $\phi$, the force can be both attractive and repulsive.
This may lead to unusual phase behaviour in smectic layers  of parallel DNA molecules.
Details of the molecular shape and counterion size are important for
small separations as well. The torque is relatively small except for
small separations where it exhibits a complicated $\phi$-dependence.

Second, as a function of added salt concentration we predict a non-monotonic
behaviour of the force induced by a competition between delocalization of
condensed counterions and enhanced electrostatic screening. This effect
 can in principle be
verified in experiments.

Third, linear screening theories describe the simulation
data qualitatively but not quantitatively. Having in mind that the total
force is dominated by the depletion term which is typically neglected in linear screening
theory, such theories need  improvement. On the other hand, the different
theories predict the correct long-distance behaviour, 
if a phenomenological fit parameter - as the renormalized phosphate charge
$q_p^*$ for the Yukawa-segment model or the condensation fraction $\theta$
for the Kornyshev-Leikin model - is introduced. The Yukawa-segment model
can  even predict the orientational dependence of the force and the torque at smaller
distances in the case of small counterion and phosphate sizes.
Hence, a phenomenological Yukawa segment  model can be used in a statistical 
description of the phase behaviour of
many parallel DNA strands in a smectic layer.

Future work should focus on an analysis for divalent counterions which are expected
to lead to a qualitatively different behaviour since the Coulomb coupling is enhanced 
strongly in this case. Also, one should step by step increase the complexity of
the model in order to take effects such as dielectric discontinuities
\cite{jaya1,maarel,stigter,skolnick}, chemical bindings
of counterions in the grooves and   discrete polarizable solvents into account.

\acknowledgments
We  thank A. A. Kornyshev, S. Leikin, G. Sutmann, H. M. Harreis, and C. N. Likos for
stimulating discussions and helpful comments. Financial
support from the Deutsche Forschungsgemeinschaft within 
the project Lo 418/6-1 
(``Theory of Interaction, recognition and assembling 
of biological helices") is gratefully acknowledged.

\twocolumn[\hsize\textwidth\columnwidth\hsize\csname
@twocolumnfalse\endcsname

\appendix
\section{Lekner summation method for forces}

In our  simulations we account for the long-range nature of the Coulomb
interactions via the efficient method proposed by Lekner
\cite{lekner}. This method has been successfully applied to  partially
periodic systems \cite{mashl,jensen}.
  For an assembly of $N$ ions in a central cubic cell of dimension $L$, the
 Coulomb force ${\vec F}_i^{(c)}$ exerted  onto particle $i$ by
 particle $j$, and by all 
repetitions of particle $j$ in the periodic system, is 
\begin{equation}
{\vec F}_i^{(c)} = \frac{q_iq_j}{\epsilon} \sum_{\rm {all \, cells}}
  \frac{\vec r_i-\vec r_j}{\vert \vec r_i-\vec r_j \vert ^3}.
\end{equation}
Because of $x,y,z$ symmetry it is sufficient to consider  only one component of
the force. For the $x$-component of the force we have
\begin{eqnarray}
{\vec F}_{ix}^{(c)}=&& \frac{q_iq_j}{\epsilon L^2} 8\pi \sum_{l=1}^{\infty}l \sin
(2\pi l \frac {\Delta x}{L})  \nonumber \\
&& \sum_{m=-\infty}^{\infty}  \sum_{n=-\infty}^{\infty} K_0 \left (2
  \pi l \left((\frac {\Delta y}{L} +m)^2+(\frac {\Delta z}{L}+n
    )^2\right)^{1/2}
 \right)
\label{leknerx}
\end{eqnarray}
Here, $\Delta x= x_i-x_j,\,\, \Delta y= y_i-y_j,\,\, \Delta z= z_i-z_j$, and  
 $K_0(z)$ is the modified Bessel function of zero order.

  For a pair of particles not aligned parallel to the $x$-axis, the convergence
of the sum in (\ref{leknerx}) is fast. Thus an evaluation of just 20 terms
in the sum is enough to get a part-per-million accuracy. The
convergence becomes worse when simultaneously $\vert \Delta y \vert < \delta$ and $
\vert \Delta z \vert < \delta$ ($\delta \ll L$) for the
case $m=0=n$. The  number of terms needed in the sum
for a desired accuracy increases rapidly with increasing $\delta$.

If the particles are aligned  parallel to the $x$-axis such
that $\vert \Delta y \vert + \vert \Delta z \vert \equiv 0$,
the sum in (\ref{leknerx}) diverges with $m=0=n$.
For this particular case $\vec F_{ix}$ is          
\begin{eqnarray}
{\vec F}_{ix}^{(c)}=&&  \frac{q_iq_j}{\epsilon L^2} \frac{8\pi}{\sqrt 2}
\sum_{l=1}^{\infty} l \sin
 (2\pi l \frac {\Delta x}{2L}) \nonumber \\
 &&\times  \sum_{m=-\infty}^{\infty} \left \lgroup
  K_0 \left (2 \pi l \vert \frac {\Delta x}{2L} +m \vert \right )
 + (-1)^l K_0 \left ( 2 \pi l \vert \frac {\Delta x}{2L} + m - {\rm
     sign} (\Delta x)\frac{1}{2} \vert \right) \right \rgroup
\label{leknerxx}
\end{eqnarray}

]

\end{document}